\documentclass[aps,prc,showpacs,showkeys,twocolumn,floatfix]{revtex4}
\usepackage[dvips]{color,graphicx}

\def\gtorder{\mathrel{\raise.3ex\hbox{$>$}\mkern-14mu
 \lower0.6ex\hbox{$\sim$}}}
\def\ltorder{\mathrel{\raise.3ex\hbox{$<$}\mkern-14mu
 \lower0.6ex\hbox{$\sim$}}}

\def\be{\begin{eqnarray}}
\def\ee{\end{eqnarray}}
\def\bea{\begin{eqnarray}}
\def\eea{\end{eqnarray}}
\def\beas{\begin{eqnarray*}}
\def\eeas{\end{eqnarray*}}
\newcommand{\eq}[1]{Eq.~(\ref{#1})}

\def\bfb{{\bf b}}

\def\bfR{{\bf R}}

\def\bfq {{\bf q}}
\def\bfv {{\bf v}}

\def\bfb{{\bf b}}
\def\bfB{{\bf B}}
\def\bfk{{\bf k}}

\def\bfp{{\bf p}}
\begin{document}

\noindent
{ NT@UW-08-12}

\title{The neutron negative central charge density: an inclusive-exclusive
connection}

\author{Gerald A. Miller}
\affiliation{Department of Physics,
University of Washington\\
Seattle, Washington 98195-1560}

\author{John Arrington}
\affiliation{Physics Division, Argonne National Laboratory\\
Argonne, Illinois 60439}

\begin{abstract}

Models of generalized parton distributions at zero skewness are used to relate
the behavior of deep inelastic scattering quark distributions, evaluated at
high $x$, to the transverse charge density evaluated at small distances.  We
obtain an interpretation of the recently obtained negative central charge
density of the neutron.  The $d$ quarks dominate the neutron structure
function for large values of Bjorken $x$, where the large momentum of the
struck quark has a significant impact on determining the center of momentum,
and thus the ``center'' of the nucleon in the transverse position plane.


\end{abstract}


\maketitle

Much experimental technique, effort and ingenuity has been used recently to
measure the electromagnetic form factors of the nucleon~\cite{gao03,
hydewright04, perdrisat07, arrington07a}.  These quantities are probability
amplitudes that the nucleon can absorb a given amount of momentum and  remain
in the ground state, and therefore should determine the nucleon charge and
magnetization densities.

In the non-relativistic case, the form factors are simply the Fourier
transforms of the rest frame spatial distributions, and the charge and
magnetization mean square radii are derived from the slope of the form factors
at $Q^2=0$.  In the relativistic case, this interpretation is not correct
because the wave functions of the initial and final  nucleons have different
momenta and therefore differ, invalidating a probability or density
interpretation. This is addressed by working in the Breit frame, where the
magnitude of the initial and final nucleon momenta are identical. However, one
needs boost corrections of order $Q^2/m^2$, where $m^2$ is the mass of the
constituent particles to which the boost is applied, to relate the rest frame
and moving nucleon wave functions.  These corrections are
model-dependent~\cite{kelly02}, so the use of the Breit frame does not provide
a precise, model independent measure of the spatial distribution of the
nucleon.

A recent work showed that it is possible to obtain a model-independent nucleon
charge density~\cite{miller07}.  In the infinite momentum frame (IMF), the
two-dimensional Fourier transform of the elastic form factor $F_1$, provides a
model-independent transverse charge distribution, $\rho_\perp(b)$, where $b$
is the distance from the center of momentum in the transverse plane.  The use
of existing data and convenient parameterizations~\cite{kelly04, bradford06}
yielded a central charge density of the neutron, $\rho_\perp^n(b=0)$, that is
negative.  We also note that the two-dimensional Fourier transform of $F_2$
can be interpreted as a magnetization density~\cite{miller07b}, and that this
yields a difference between the magnetic and electric radii in the proton.

These findings appear to contradict previous understanding of the nucleon
charge and magnetization distributions based on the model-dependent extraction
of the rest frame charge distributions.  The negative core of the neutron
transverse density also contradicts previous intuition that the component in
which the neutron is represented as a proton surrounded by a negatively
charged pion cloud causes the central charge density to be positive.  This
negative core is a feature even in models that  include  a pion cloud effect
to reproduce the measured values of $F_1^n$.   It is therefore important to
understand the differences between this model-independent transverse charge
density and the rest frame charge density to fully understand the new features
of the transverse spatial distributions.

Our goal is to obtain further information about the neutron charge
density by using generalized parton distributions (GPDs) which contain
information about the longitudinal momentum fraction $x$ as well as the
transverse position $b$.  Experimental information regarding the $x$
dependence is obtained by using GPDs to reproduce both deep inelastic 
scattering and elastic scattering data. Thus we use this inclusive-exclusive
connection to better understand the central neutron charge density.

To start the analysis, we recall that form factors are matrix elements of the
electromagnetic current operator $J^\mu(x^\nu)$ in units of the proton charge.
The momentum transfer $q$ is space-like, so that $Q^2 \equiv -q^2 >
0$. The normalization is such that  $F_1(0)$ is the nucleon charge, and 
$F_2(0)$ is the proton anomalous magnetic moment.  The Sachs electric
and magnetic form factors are given by $G_E=F_1-(Q^2/4M^2) F_2$ and
$G_M=F_1+F_2$.

The widely studied GPDs \cite{ji96, radyushkin97} are of high current interest
because they can be related to the total angular momentum carried by quarks in
the nucleon. We consider the specific case in which the longitudinal momentum
transfer $\xi$ is zero, and the initial and final nucleon helicities are
identical ($\lambda'=\lambda$). Then, in the light-cone gauge, $A^+=0$, the
matrix element defining the GPD $H_q$ for a quark of flavor $q$ and zero
skewness is
\be
H_q(x,t) = \int\!\! \frac{dx^-}{4\pi}\langle p^+,\bfp',\lambda|\widehat{O}_q(x,{\bf 0})
|p^+,\bfp,\lambda\rangle e^{ixp^+x^-},
\label{eq:pd}
\eea
where
\be
\widehat{O}_q(x,{\bf b}) \equiv
\int \frac{dx^-}{4\pi}{q}_+^\dagger
\left(-\frac{x^-}{2},{\bf b} \right) 
q_+\left(\frac{x^-}{2},{\bf b}\right) 
e^{ixp^+x^-}.
\label{eq:bperp}
\ee
We abbreviate $H_q(x,\xi$=$0,t)\equiv H_q(x,t)$ and $
-t=-(p'-p)^2=(\bfp'-\bfp)^2=Q^2.$ The simple form of $t$ results from its
invariance under transverse boosts~\cite{kogut70}: Lorentz transformations,
defined by a transverse vector $\bfv$  that transform a four-vector $k$
according to $k^+\rightarrow k^+,\;\bfk\rightarrow\bfk-k^+\bfv$ and $k^-$ such
that $k^2$ is unchanged.  These quantities are part of a kinematic subgroup of
the Poincar\'{e} group that obey the same commutation relations as those among
the generators of the Galilean transformations for non-relativistic quantum
mechanics in the transverse plane. 

GPDs allow for a unified description of a number of hadronic properties
\cite{ji96}. The most relevant for us are that for $t$=0 they reduce to
conventional PDFs, $H_q(x,0)= q(x)$, and that the integration of the
charge-weighted $H_q$ over $x$ yields the nucleon electromagnetic form factor:
\be
F_1(t)=\sum_q e_q \int dx H_q(x,t).  
\label{eq:form}
\ee

The spatial structure of a  nucleon can be examined~\cite{soper77, burkardt03,
diehl02,burkardt00} using nucleonic states that are transversely localized. 
The state with transverse center of mass $\bfR$ set to 0, $\left| p^+,{\bf
R}={\bf 0}, \lambda\right\rangle$ is formed by taking a linear superposition
of states of transverse momentum. Doing this requires the use of a frame with
infinitely large $p^+$.

The impact parameter-dependent PDF~\cite{burkardt00} is the matrix element
of the operator $\widehat {O}_q$ in the state $\left|p^+,{\bf R}= {\bf 0},
\lambda\right\rangle$:
\be
\rho_\perp^q({\bf b},x) \equiv 
\left\langle p^+,{\bf R}= {\bf 0},
\lambda\right|
\widehat{O}_q(x,{\bf b})
\left|p^+,{\bf R}= {\bf 0},
\lambda\right\rangle. 
\label{eq:def1}
\ee
We use the notation $\rho_\perp^q({\bf b},x)$ instead of the originally
defined~\cite{burkardt00} $q(x,\bfb)$ because the quantity truly is a
density,  giving the probability that the quark has  a longitudinal momentum
fraction $x$ and is at a transverse position $\bfb$.
The quantity $\rho_\perp^q({\bf b},x)$ is the two-dimensional Fourier
transform of the GPD $H_q$:
\bea
\rho_\perp^q({\bf  b},x)=\int {d^2q\over
(2\pi)^2}e^{-i\;\bfq\cdot\bfb}H_q(x,t=-\bfq^2),\label{ft1}
\eea
with $H_q$ appearing because the initial and final helicities are each
$\lambda$. A complete determination of $H_q(x,t)$ (with $t\le0$) would 
determine $\rho_\perp(x,{\bf b})$.

One finds can extract the form factor $F_1$~\cite{soper77} by integrating
$\rho_\perp^q({\bf b},x)$ over all values of $x$, multiplying by the quark
charge $e_q$, and summing over quark flavors $q$. The resulting IMF charge
density in transverse space is
\bea
&\rho_\perp^N(b) & \equiv \sum_q e_q\int dx\;\rho_\perp^q({\bf b},x) \cr
&&=\int {d^2q\over (2\pi)^2} F_1(Q^2=\bfq^2) e^{-i\;\bfq\cdot\bfb}.
\label{rhobt}
\eea
This quantity gives the charge density at a transverse position $\bfb$
irrespective of the longitudinal momentum fraction. The primary difference
between the present  charge density $\rho_\perp(b)$ and the older
interpretation that the charge density as the three-dimensional Fourier
transform of $G_E$ is that the present approach provides a model-independent,
two-dimensional charge distribution in the plane transverse to the motion of
the nucleon in the infinite momentum frame.  The boost corrections here are
simply kinematic and are incorporated in the formalism. In the older
interpretation, the model-dependent boost corrections can not be avoided.

Our aim is to investigate $\rho_\perp(\bfb,x)$ to understand the origin of the
neutron's negative central charge density. The quantities are not measured
directly, but have been obtained from models that incorporate fits to parton
distributions and electromagnetic nucleon form factors~\cite{guidal05,
diehl05, ahmad07,tiburzi04}. This method exploits form factor sum rules at zero
skewness, obtained neglecting the effects of strangeness, to obtain
information regarding the valence quark GPDs, $H^q_v\equiv H^q-H^{\bar{q}}$.
This yields the net contribution to the form factors from quarks and
anti-quarks, although it does  not correspond to the valence distribution
within a model for which   sea distributions for quarks and antiquarks have
different $x$ or $t$ dependences. To proceed further one must model the GPDs,
and the results can be expected to depend on the chosen forms. Diehl {\it et
al.}~\cite{diehl05} use
\bea
H^q_v(x,t)=q_v(x) \exp[f_q(x)t],
\eea
where 
\bea
f_q(x)=[\alpha'\log[1/x]+B_q](1-x)^3+A_qx(1-x)^2,
\eea
is the form that gives the best fit to the data. The parameter $\alpha'$
represents the slope of the Regge trajectory ($\alpha'=0.9$  GeV$^2$), and the
CTEQ6 PDFs~\cite{pumplin02} are taken as input. Here we use the best fit
parameters, taken from the second line of Table 8 of~\cite{diehl05}. These are
$A_u=1.26\;{\rm GeV}^{-2}, B_u=0.59\;{\rm GeV}^{-2}, A_d=3.82\;{\rm GeV}^{-2},
B_d=0.32\;{\rm GeV}^{-2}.$ We note that the labels $u$ and $d$ here refer to
the $u$ and $d$ quarks in the proton. These correspond to $d$ and $u$ quarks
in the neutron, if charge symmetry~\cite{miller90, miller98, londergan98,
miller06} is upheld. It is well-known that for the proton, $2d_v/u_v$ falls
rapidly for large values of $x$, which means that $u$ quarks dominate the
parton distribution for large  values of $x$. This means that in the neutron,
the $d$ quarks dominate the parton distribution for large values of $x$. The
distributions of~\cite{guidal05} have $A_q=B_q=0$ and
$f_q(x)=[\alpha'_q\log[1/x]](1-x).$ Those of~\cite{ahmad07} have a more
complicated form and also include the constraint that the nucleon consists of
three quarks at an initial scale of $Q_0^2=0.094$ GeV$^2$.

\begin{center}
 \begin{figure}[htb]
   \includegraphics[width=8.0cm,height=4.5cm]{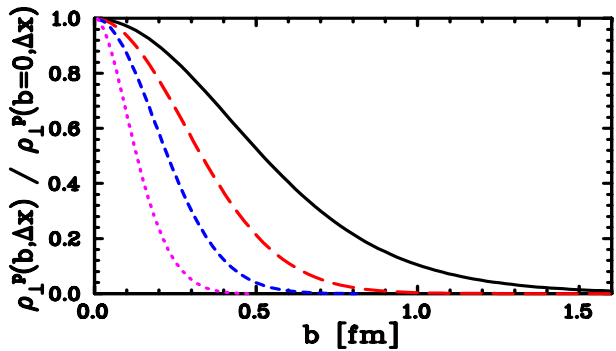}
   \caption{(Color Online)  
The proton transverse charge density, $\rho_\perp^p(b,\Delta x)$, for quarks in
different $\Delta x$ regions: $x$$<$0.15 (solid), 0.15$<$$x$$<$0.3 (long-dash),
0.3$<$$x$$<$0.5 (short-dash), and $x$$>$0.5 (dotted).  The curves are
calculated from the GPD of Ref.~\cite{diehl05} and have been
normalized to unity at $b=0$.}
     \label{fig:proton_scaled}
 \end{figure}
\end{center}

Our goal here is to examine the connection between regions of $x$ and regions
of $b$. To do this we define
\bea
\rho_\perp^q(b,\Delta x)\equiv \int_{\Delta x} dx~e_q~\rho_\perp^q(b,x),
\eea
with $\rho_\perp^{p,n}$ being obtained from
appropriate sums of $\rho_\perp^{q}$.  This represents the contribution to the
charge density from quarks in the $x$ region defined by $\Delta x$.

An important feature of the present approach is that these charge
distributions are taken with respect to the center of momentum in the
transverse plane.  Thus, the transverse position $b$ is taken with respect to
the  momentum-weighted average position of all partons, including the struck
quark.  At low $x$, the struck  quark has little impact on the center of
momentum, and this corresponds to intuitive picture of spatial distribution.
At large $x$, the struck quark plays a significant role in defining the CM,
and so distribution becomes localized at small values of $b$.  This can be
seen in Fig.~\ref{fig:proton_scaled}, where for $x \approx 0.1$,
the half-maximum width is 0.5~fm, while for $x \approx 0.8$, it is 0.12~fm. 
The curves have been scaled to yield unity at $b=0$, to emphasize the
variation in width.  The four $\Delta x$ regions yield 58\%, 25\%, 14\%, and
3\% of the total charge, with the largest contributions coming from the bins
with the  smallest values of  $x$. Thus the large $x$ quarks, dominantly $u$
quarks in the proton, play an increasingly prominent role in the charge
distribution at small values of $b$. The figure obtained using the Guidal
\textit{et al}, parameterization for the GPDs is barely distinguishable from
Fig.~\ref{fig:proton_scaled}.  The GPDs of \cite{ahmad07} also have a strong
tendency to be constrained to smaller and smaller values of $b$ as the value
of $x$ increases.  We evaluate the GPDs of all three models using the
starting scale $Q_0^2$ of each model.

\begin{center}
 \begin{figure}[htb]
   \includegraphics[width=8.0cm,height=6cm]{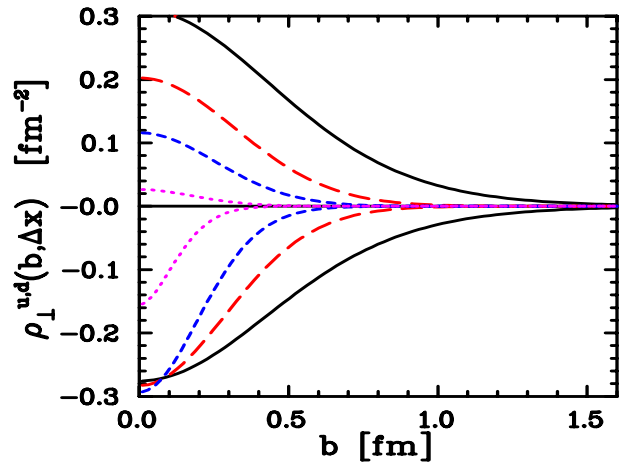}
   \caption{(Color Online) The $u$ and $d$ quark contributions to the neutron
transverse charge density, $\rho_\perp^u(b,\Delta x)$ and
$\rho_\perp^d(b,\Delta x)$. The curves correspond to the same $\Delta x$
regions as in Fig.~\ref{fig:proton_scaled}. The largest contributions come
from small $x$, where $u$ and $d$ quarks contribute roughly equal
amounts of charge. As one goes to larger $x$ values, the charge is shifted to
smaller values of $b$, while at the same time the contribution from the up
quarks drops rapidly with respect to the down quarks, due to the rapid falloff
of the neutron $u$ to $d$ quark ratio at large $x$.}
   \label{fig:neutron_bins}
 \end{figure}
\end{center}

Now consider the charge distribution of the neutron. We expect that the $d$
quarks dominate at large $x$ and therefore become important at small values of
$b$. Because the distribution of quarks at large $x$ will be highly localized
near $b=0$, a negative peak can be formed if the large $x$ distribution is
sufficiently dominated by down quarks, thus yielding a significant
contribution of negative charge at large enough $x$. At very low $x$ values,
the valence distribution for up quarks in the neutron is roughly half that of
the down quarks, $d_v^n(x)/u_v^n(x) \approx 2$, and the net charge coming from
$u$ and $d$ quarks will approximately cancel, although the distribution as a
function of $b$ need not be zero everywhere. Above $x=0.5$, $d_v^n(x)$ is at
least three times the size of $u_v^n(x)$, and increases with $x$.  So
for $x>0.5$ the net impact to the charge distribution will be negative, and
will be peaked at smaller values of $b$.  We show this explicitly in
Fig.~\ref{fig:neutron_bins}, where we separate the contributions to the
neutron charge density from $u$ and $d$ quarks based on the GPD
fit of Ref.~\cite{diehl05}.  The distributions of~\cite{guidal05}
and~\cite{ahmad07} yield somewhat different results, but they exhibit the same
qualitative behavior.

\begin{center}
 \begin{figure}[htb]
   \includegraphics[width=8.0cm,height=4.5cm]{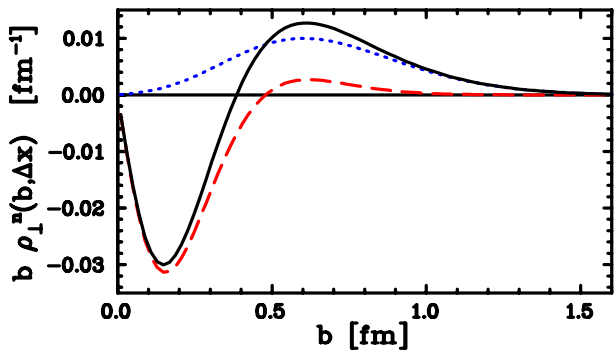}
   \caption{(Color online) Transverse charge density for the neutron.  The
dotted line is the contribution from $x<0.23$, dashed is that for $x>0.23$, and
the solid is the total.}
   \label{fig:neutron_total}
 \end{figure}
\end{center}

The next step is to examine the total charge distribution of the neutron.
Fig.~\ref{fig:neutron_total} separates the contributions from low and high $x$
regions. For $x<0.23$ the charge distribution is positive for $b<1.5$~fm and
slightly negative distribution at larger radii.  For $x>0.23$, the
contribution is largely negative, and highly localized below 0.5~fm. The
negative region at the center of the neutron transverse charge distribution
arises a natural consequence of the model-independent definition of the charge
density.  The low momentum partons have a larger spatial extent and reproduce
the intuitive result of the pion cloud picture: a positive core with a small
negative tail at large distances.

\begin{center}
 \begin{figure}[htb]
   \includegraphics[width=8.0cm,height=5.5cm]{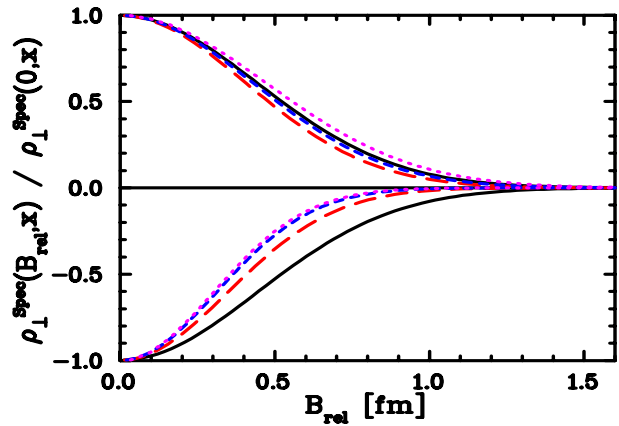}
   \caption{(Color online) The $u$ and $d$ quark contributions to 
$\rho^{{\rm Spec},n}_\perp(B_{\rm rel},x)$ see \eq{redefine}. vs $B_{\rm rel}$
for $x=0.1$~(solid), 0.3~(long-dash), 0.5~(short-dash), and 0.7~(dotted). The
curves are scaled to unity at $B_{\rm rel}=0$.}
   \label{fig:neutron_spectator}
 \end{figure}
\end{center}

A more intuitive picture of the charge distribution can be obtained by
looking at the distribution of charge relative to the spectator partons, so
that the struck quark does not influence the definition of the center of mass.
This can be approximated by looking at the position of the struck quark
relative to the spectators.  We work in the transverse plane, with the
origin set to the center of momentum, giving $\sum_{i}x_i{\bf b_i}=0$. For a
struck quark at $(x_1,{\bf b_1})\equiv(x,\bfb)$, we can determine the
momentum-weighted spectator position, ${\bf b_{\rm spec}}$, and the relative
distance from the struck quark to the spectator quarks:
\bea
x_1 \bfb_1+\sum_{i>1}x_i\bfb_i=x\bfb+(1-x)\bfb_{\rm spec}=0,\\
{\bf B_{\rm rel}} = \bfb - {\bf b_{\rm spec}}=\frac{\bfb}{(1-x)}=\bf{B_{\rm rel}}.
\eea
We exhibit the dependence on ${\bf B_{\rm rel}}$ by defining a function
\bea
\rho^{\rm Spec}_\perp({\bf B}_{rel},x)\equiv \rho_\perp(\bfB_{rel}(1-x),x)\label{redefine}
\eea
which gives the probability that a struck quark of longitudinal momentum
fraction $x$ is a distance ${\bf B_{\rm rel}}$ away from the spectator center of
momentum.  Figure~\ref{fig:neutron_spectator} shows this rescaled version of
$\rho_\perp(b)$, with the contribution at each $x$ value normalized to
unity at $b=0$.  
The quantity $\rho^{\rm Spec}_\perp({\bf B}_{\rm rel},x)$ can not be determined in
a model independent manner,  but
may be a better approximation to our intuitive picture of the charge
distribution, as it removes the influence of the struck quark on defining the
center of the nucleon. While the charge distribution coming from very low $x$
quarks has a greater spatial extent, the decreasing width of the
$\rho_\perp(b)$ distribution for large $x$ quarks is essentially completely
removed when looking at $B_{\rm rel}$.

Before concluding, it is worthwhile to comment on the relation between the
present work and the difference between the electric and magnetic radii of the
proton~\cite{miller07b}.  In the model-independent, IMF approach presented
here, the electric and magnetic transverse radii have a clear connection to
$F_1$ and $F_2$ and a Foldy~\cite{foldy51} term causes a difference 
between the transverse radii. The Foldy term is responsible for most of the
charge radius defined by $G_E$.  Understanding the neutron's negative central
density is more subtle and requires knowledge of $\rho(x,\bfb)$.

We summarize our findings with the statement that, using the model GPDs of
Refs.~\cite{diehl05, guidal05, ahmad07}, the dominance of the neutron's $d$
quarks at high values of $x$ leads to a negative contribution to the charge
density which, due to the definition of $\bfb$, becomes localized near the
center of mass of the neutron.  This localization does not appear when
examined as a function of the position of the struck quark relative to the
spectators.

\begin{acknowledgments}

This work was supported by the U. S. Department of Energy, Office of Nuclear
Physics, under contracts FG02-97ER41014 and DE-AC02-06CH11357.  We thank
D. Geesaman, P.~Kroll and B.~Wojtsekhowski for useful discussions. We thank the ECT* for
hosting a workshop where many  of the calculations we present were performed.

\end{acknowledgments}

\bibliography{interpretation}

\begin{thebibliography}{26}
\expandafter\ifx\csname natexlab\endcsname\relax\def\natexlab#1{#1}\fi
\expandafter\ifx\csname bibnamefont\endcsname\relax
  \def\bibnamefont#1{#1}\fi
\expandafter\ifx\csname bibfnamefont\endcsname\relax
  \def\bibfnamefont#1{#1}\fi
\expandafter\ifx\csname citenamefont\endcsname\relax
  \def\citenamefont#1{#1}\fi
\expandafter\ifx\csname url\endcsname\relax
  \def\url#1{\texttt{#1}}\fi
\expandafter\ifx\csname urlprefix\endcsname\relax\def\urlprefix{URL }\fi
\providecommand{\bibinfo}[2]{#2}
\providecommand{\eprint}[2][]{\url{#2}}

\bibitem[{\citenamefont{Gao}(2003 [Erratum-ibid., 567, 2003])}]{gao03}
\bibinfo{author}{\bibfnamefont{H.}~\bibnamefont{Gao}}, \bibinfo{journal}{Int.
  J. Mod. Phys.} \textbf{\bibinfo{volume}{E12}}, \bibinfo{pages}{1}
  (\bibinfo{year}{2003 [Erratum-ibid., 567, 2003]}).

\bibitem[{\citenamefont{Hyde-Wright and de~Jager}(2004)}]{hydewright04}
\bibinfo{author}{\bibfnamefont{C.~E.} \bibnamefont{Hyde-Wright}}
  \bibnamefont{and} \bibinfo{author}{\bibfnamefont{K.}~\bibnamefont{de~Jager}},
  \bibinfo{journal}{Ann. Rev. Nucl. Part. Sci.} \textbf{\bibinfo{volume}{54}},
  \bibinfo{pages}{217} (\bibinfo{year}{2004}).

\bibitem[{\citenamefont{Perdrisat et~al.}(2007)\citenamefont{Perdrisat,
  Punjabi, and Vanderhaeghen}}]{perdrisat07}
\bibinfo{author}{\bibfnamefont{C.~F.} \bibnamefont{Perdrisat}},
  \bibinfo{author}{\bibfnamefont{V.}~\bibnamefont{Punjabi}}, \bibnamefont{and}
  \bibinfo{author}{\bibfnamefont{M.}~\bibnamefont{Vanderhaeghen}},
  \bibinfo{journal}{Prog. Part. Nucl. Phys.} \textbf{\bibinfo{volume}{59}},
  \bibinfo{pages}{694} (\bibinfo{year}{2007}).

\bibitem[{\citenamefont{Arrington et~al.}(2007)\citenamefont{Arrington,
  Roberts, and Zanotti}}]{arrington07a}
\bibinfo{author}{\bibfnamefont{J.}~\bibnamefont{Arrington}},
  \bibinfo{author}{\bibfnamefont{C.~D.} \bibnamefont{Roberts}},
  \bibnamefont{and} \bibinfo{author}{\bibfnamefont{J.~M.}
  \bibnamefont{Zanotti}}, \bibinfo{journal}{J. Phys.}
  \textbf{\bibinfo{volume}{G34}}, \bibinfo{pages}{S23} (\bibinfo{year}{2007}).

\bibitem[{\citenamefont{Kelly}(2002)}]{kelly02}
\bibinfo{author}{\bibfnamefont{J.~J.} \bibnamefont{Kelly}},
  \bibinfo{journal}{Phys. Rev. C} \textbf{\bibinfo{volume}{66}},
  \bibinfo{pages}{065203} (\bibinfo{year}{2002}).

\bibitem[{\citenamefont{Miller}(2007)}]{miller07}
\bibinfo{author}{\bibfnamefont{G.~A.} \bibnamefont{Miller}},
  \bibinfo{journal}{Phys. Rev. Lett.} \textbf{\bibinfo{volume}{99}},
  \bibinfo{pages}{112001} (\bibinfo{year}{2007}).

\bibitem[{\citenamefont{Kelly}(2004)}]{kelly04}
\bibinfo{author}{\bibfnamefont{J.~J.} \bibnamefont{Kelly}},
  \bibinfo{journal}{Phys. Rev. C} \textbf{\bibinfo{volume}{70}},
  \bibinfo{pages}{068202} (\bibinfo{year}{2004}).

\bibitem[{\citenamefont{Bradford et~al.}(2006)\citenamefont{Bradford, Bodek,
  Budd, and Arrington}}]{bradford06}
\bibinfo{author}{\bibfnamefont{R.}~\bibnamefont{Bradford}},
  \bibinfo{author}{\bibfnamefont{A.}~\bibnamefont{Bodek}},
  \bibinfo{author}{\bibfnamefont{H.}~\bibnamefont{Budd}}, \bibnamefont{and}
  \bibinfo{author}{\bibfnamefont{J.}~\bibnamefont{Arrington}},
  \bibinfo{journal}{Nucl. Phys. Proc. Suppl.} \textbf{\bibinfo{volume}{159}},
  \bibinfo{pages}{127} (\bibinfo{year}{2006}).

\bibitem[{\citenamefont{Miller et~al.}(2007)\citenamefont{Miller, Piasetzky,
  and Ron}}]{miller07b}
\bibinfo{author}{\bibfnamefont{G.~A.} \bibnamefont{Miller}},
  \bibinfo{author}{\bibfnamefont{E.}~\bibnamefont{Piasetzky}},
  \bibnamefont{and} \bibinfo{author}{\bibfnamefont{G.}~\bibnamefont{Ron}}
  (\bibinfo{year}{2007}), \eprint{0711.0972}.

\bibitem[{\citenamefont{Ji}(1997)}]{ji96}
\bibinfo{author}{\bibfnamefont{X.-D.} \bibnamefont{Ji}},
  \bibinfo{journal}{Phys. Rev.} \textbf{\bibinfo{volume}{D55}},
  \bibinfo{pages}{7114} (\bibinfo{year}{1997}), \eprint{hep-ph/9609381}.

\bibitem[{\citenamefont{Radyushkin}(1997)}]{radyushkin97}
\bibinfo{author}{\bibfnamefont{A.~V.} \bibnamefont{Radyushkin}},
  \bibinfo{journal}{Phys. Rev.} \textbf{\bibinfo{volume}{D56}},
  \bibinfo{pages}{5524} (\bibinfo{year}{1997}), \eprint{hep-ph/9704207}.

\bibitem[{\citenamefont{Kogut and Soper}(1970)}]{kogut70}
\bibinfo{author}{\bibfnamefont{J.~B.} \bibnamefont{Kogut}} \bibnamefont{and}
  \bibinfo{author}{\bibfnamefont{D.~E.} \bibnamefont{Soper}},
  \bibinfo{journal}{Phys. Rev. D} \textbf{\bibinfo{volume}{1}},
  \bibinfo{pages}{2901} (\bibinfo{year}{1970}).

\bibitem[{\citenamefont{Diehl}(2002)}]{diehl02}
\bibinfo{author}{\bibfnamefont{M.}~\bibnamefont{Diehl}}, \bibinfo{journal}{Eur.
  Phys. J.} \textbf{\bibinfo{volume}{C25}}, \bibinfo{pages}{223}
  (\bibinfo{year}{2002}).

\bibitem[{\citenamefont{Soper}(1977)}]{soper77}
\bibinfo{author}{\bibfnamefont{D.~E.} \bibnamefont{Soper}},
  \bibinfo{journal}{Phys. Rev. D} \textbf{\bibinfo{volume}{15}},
  \bibinfo{pages}{1141} (\bibinfo{year}{1977}).

\bibitem[{\citenamefont{Burkardt}(2003)}]{burkardt03}
\bibinfo{author}{\bibfnamefont{M.}~\bibnamefont{Burkardt}},
  \bibinfo{journal}{Int. J. Mod. Phys.} \textbf{\bibinfo{volume}{A18}},
  \bibinfo{pages}{173} (\bibinfo{year}{2003}).

\bibitem[{\citenamefont{Burkardt}(2000)}]{burkardt00}
\bibinfo{author}{\bibfnamefont{M.}~\bibnamefont{Burkardt}},
  \bibinfo{journal}{Phys. Rev. D} \textbf{\bibinfo{volume}{62}},
  \bibinfo{pages}{071503(R)} (\bibinfo{year}{2000}).

\bibitem[{\citenamefont{Diehl et~al.}(2005)\citenamefont{Diehl, Feldmann, and
  Kroll}}]{diehl05}
\bibinfo{author}{\bibfnamefont{M.}~\bibnamefont{Diehl}},
  \bibinfo{author}{\bibfnamefont{R.}~\bibnamefont{Feldmann}, \bibfnamefont{Th.
  an~Jakob}}, \bibnamefont{and}
  \bibinfo{author}{\bibfnamefont{P.}~\bibnamefont{Kroll}},
  \bibinfo{journal}{Eur. Phys. J.} \textbf{\bibinfo{volume}{C39}},
  \bibinfo{pages}{1} (\bibinfo{year}{2005}).

\bibitem[{\citenamefont{Guidal et~al.}(2005)\citenamefont{Guidal, Polyakov,
  Radyushkin, and Vanderhaeghen}}]{guidal05}
\bibinfo{author}{\bibfnamefont{M.}~\bibnamefont{Guidal}},
  \bibinfo{author}{\bibfnamefont{M.~V.} \bibnamefont{Polyakov}},
  \bibinfo{author}{\bibfnamefont{A.~V.} \bibnamefont{Radyushkin}},
  \bibnamefont{and}
  \bibinfo{author}{\bibfnamefont{M.}~\bibnamefont{Vanderhaeghen}},
  \bibinfo{journal}{Phys. Rev. D} \textbf{\bibinfo{volume}{72}},
  \bibinfo{pages}{054013} (\bibinfo{year}{2005}).

\bibitem[{\citenamefont{Ahmad et~al.}(2007)}]{ahmad07}
\bibinfo{author}{\bibfnamefont{S.}~\bibnamefont{Ahmad}} \bibnamefont{et~al.},
  \bibinfo{journal}{Phys. Rev. D} \textbf{\bibinfo{volume}{75}},
  \bibinfo{pages}{094003} (\bibinfo{year}{2007}).

\bibitem[{\citenamefont{Tiburzi et~al.}(2004)\citenamefont{Tiburzi, Detmold,
  and Miller}}]{tiburzi04}
\bibinfo{author}{\bibfnamefont{B.~C.} \bibnamefont{Tiburzi}},
  \bibinfo{author}{\bibfnamefont{W.}~\bibnamefont{Detmold}}, \bibnamefont{and}
  \bibinfo{author}{\bibfnamefont{G.~A.} \bibnamefont{Miller}},
  \bibinfo{journal}{Phys. Rev.} \textbf{\bibinfo{volume}{D70}},
  \bibinfo{pages}{093008} (\bibinfo{year}{2004}), \eprint{hep-ph/0408365}.

\bibitem[{\citenamefont{Pumplin et~al.}(2002)}]{pumplin02}
\bibinfo{author}{\bibfnamefont{J.}~\bibnamefont{Pumplin}} \bibnamefont{et~al.},
  \bibinfo{journal}{JHEP} \textbf{\bibinfo{volume}{07}}, \bibinfo{pages}{012}
  (\bibinfo{year}{2002}).

\bibitem[{\citenamefont{Miller et~al.}(1990)\citenamefont{Miller, Nefkens, and
  Slaus}}]{miller90}
\bibinfo{author}{\bibfnamefont{G.~A.} \bibnamefont{Miller}},
  \bibinfo{author}{\bibfnamefont{B.~M.~K.} \bibnamefont{Nefkens}},
  \bibnamefont{and} \bibinfo{author}{\bibfnamefont{I.}~\bibnamefont{Slaus}},
  \bibinfo{journal}{Physics Reports} \textbf{\bibinfo{volume}{194}},
  \bibinfo{pages}{1} (\bibinfo{year}{1990}).

\bibitem[{\citenamefont{Miller}(1998)}]{miller98}
\bibinfo{author}{\bibfnamefont{G.~A.} \bibnamefont{Miller}},
  \bibinfo{journal}{Phys. Rev. C} \textbf{\bibinfo{volume}{57}},
  \bibinfo{pages}{1492} (\bibinfo{year}{1998}).

\bibitem[{\citenamefont{Londergan and Thomas}(1998)}]{londergan98}
\bibinfo{author}{\bibfnamefont{J.~T.} \bibnamefont{Londergan}}
  \bibnamefont{and} \bibinfo{author}{\bibfnamefont{A.~W.}
  \bibnamefont{Thomas}}, \bibinfo{journal}{Prog. Part. Nucl. Phys.}
  \textbf{\bibinfo{volume}{41}}, \bibinfo{pages}{49} (\bibinfo{year}{1998}),
  \eprint{hep-ph/9806510}.

\bibitem[{\citenamefont{Miller et~al.}(2006)\citenamefont{Miller, Opper, and
  Stephenson}}]{miller06}
\bibinfo{author}{\bibfnamefont{G.~A.} \bibnamefont{Miller}},
  \bibinfo{author}{\bibfnamefont{A.~K.} \bibnamefont{Opper}}, \bibnamefont{and}
  \bibinfo{author}{\bibfnamefont{E.~J.} \bibnamefont{Stephenson}},
  \bibinfo{journal}{Ann. Rev. Nucl. Part. Sci.} \textbf{\bibinfo{volume}{56}},
  \bibinfo{pages}{253} (\bibinfo{year}{2006}).

\bibitem[{\citenamefont{Foldy}(1951)}]{foldy51}
\bibinfo{author}{\bibfnamefont{L.~L.} \bibnamefont{Foldy}},
  \bibinfo{journal}{Phys. Rev.} \textbf{\bibinfo{volume}{83}},
  \bibinfo{pages}{688} (\bibinfo{year}{1951}).

\end{thebibliography}

\end{document}